# Hardware Accelerated Power Estimation

Joel Coburn, Srivaths Ravi, and Anand Raghunathan
NEC Laboratories America, 4 Independence Way, Princeton, NJ 08540
{jcoburn,sravi,anand}@nec-labs.com

## Abstract

In this paper, we present **power emulation**, a novel design paradigm that utilizes hardware acceleration for the purpose of fast power estimation. Power emulation is based on the observation that the functions necessary for power estimation (power model evaluation, aggregation, *etc.*) can be implemented as hardware circuits. Therefore, we can enhance any given design with "power estimation hardware", map it to a prototyping platform, and exercise it with any given test stimuli to obtain power consumption estimates. Our empirical studies with industrial designs reveal that power emulation can achieve significant speedups (10X to 500X) over state-of-the-art commercial register-transfer level (RTL) power estimation tools.

## 1 INTRODUCTION

Power estimation is an important part of the design process for most integrated circuits. The problem of power estimation has been researched at varying levels of abstraction, and many of the resulting techniques have been incorporated into commercial power estimation tools. However, efficient power estimation for large designs remains a challenge due to increases in design and testbench complexity. For example, RTL power estimation for a 1.25 million transistor MPEG4 decoder circuit (for an input stimulus containing 4 frames of a video stream) required 43 minutes and 55 minutes to run on two different state-of-the-art commercial tools [1, 2]. Power estimation tools that operate at the transistor and gate levels are known to be much (10X to 100X) slower.

To achieve efficient power estimation, we harness the speedup of hardware acceleration that comes from using an emulation platform. Typically, power estimation is dominated by the evaluation of component power models based on circuit inputs during simulation. The key observation of this work is that power model functionality can be implemented as hardware circuits. In practice, we can augment any design with power estimation hardware and map it to a hardware emulation platform such as an FPGA. *To the best of our knowledge, this is the first effort to leverage hardware acceleration for the purpose of power estimation.* We apply the concept of power emulation to power estimation at the register-transfer level, and present a design flow incorporating this idea. We demonstrate the benefits of power emulation by reporting the speedups in power estimation time for several large designs. The results indicate that power emulation can significantly enhance the scope of current RTL and gate-level power estimation methods by making them applicable to large designs with little or no tradeoff in accuracy. Much like functional emulation, we believe that power emulation will provide designers with the ability to study the power consumption of a design under realistic environments and operating conditions.

### 1.1 Related Work

Power estimation techniques have been explored extensively at the transistor, gate, and register-transfer levels of design abstraction [3, 4, 5, 6]. At the transistor level, power estimation is usually performed as a by-product of circuit simulation. For gate-level power estimation, signal statistics for the signals in the circuit can be computed through simulation, probabilistic analysis, or simulation with statistical sampling [4, 5]. Of these, simulation with a comprehensive testbench is the most commonly used technique, due to its accuracy and the ability to produce detailed feedback on power consumption. At the register-transfer level, approaches to power estimation include analytical techniques [7], characterization-based macromodels [8], and fast synthesis into gate-level descriptions [9]. Behavioral descriptions of hardware lack circuit structural information, so hardware power estimation at the behavioral level has limited accuracy. At the system level, most research has focused on developing power models for different system components, including processors, memories, on-chip buses, *etc.*

With rising circuit complexities, current power estimation tools perform poorly for large designs and can often be applied to only small parts of a design. The proposed idea of leveraging emulation platforms for power estimation is complementary to most previous work on power estimation, and can be applied to any simulation-based power estimation technique in which the power models may be expressed as synthesizable functions.

## 2 POWER EMULATION

The basic concept of power emulation is applicable at multiple levels of abstraction. However, in this paper, we discuss power emulation in the context of register-transfer level (RTL) power estimation. We describe RTL power estimation for an example design, and explain how the power estimation functions can be expressed as hardware and mapped to an emulation platform. We then present a design flow for power emulation. Finally, we provide experimental results to show the speedups achievable through power emulation.

### 2.1 Power Estimation Hardware

Let us consider a characterization-based power estimation methodology [6]. Power estimation is based on a "power macromodel library" for a universal set of RTL components that is created by characterizing their gate- or transistor-level implementations. For each RTL component in the circuit under consideration, a power macromodel is used to compute the component's power consumption as a function of the values observed at its input/output signals during simulation.

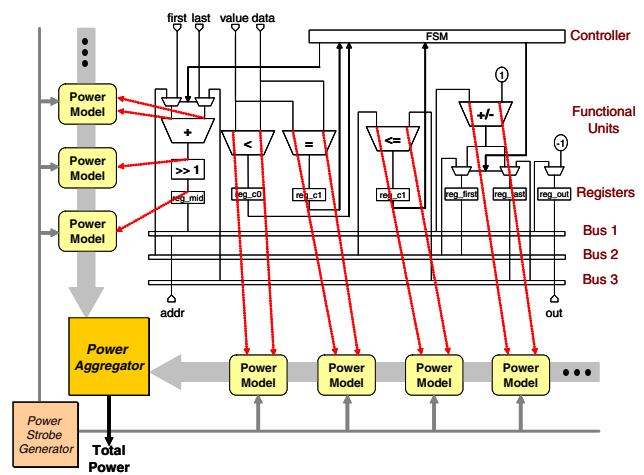

Figure 1: An example RTL circuit enhanced for power emulation

Power emulation is performed by enhancing a circuit with the appropriate power estimation hardware. This can be done by recognizing the functions pertinent to power estimation, expressing them as synthesizable circuit descriptions, and integrating them into the design. To illustrate this process, Fig. 1 shows an example RTL circuit used to perform binary search, which has been enhanced for power emulation. The circuit contains several new components for power estimation (shown shaded in Fig. 1), including (i) *power models* that are instantiated for every RTL component in order to track the component's input/output signals and compute a power value whenever triggered, (ii) a *power strobe generator* to trigger the evaluation of the power models, and (iii) a *power aggregator* to accumulate the power consumption of individual RTL components to compute the circuit's total power consumption.

Power strobe generation is done separately for each clock domain in the design. Power aggregation is simply implemented as a sequence of additions to accumulate the outputs of the power models. We focus our attention on power models, which form the most significant part of the power estimation hardware.

To explain the structure of hardware power models, we consider a cycle-accurate linear regression based macromodel [8], which expresses the power consumed in an RTL component with $n$ input/output bits as $\sum_{i=1}^{n} Coeff_i * T(x_i)$, where $Coeff_i$ represent the power model coefficients, and $T(x_i)$ is the transition count (0 or 1) at each input/output bit. The inputs to a power model include the input/output bits of the component being monitored, and a power strobe. The output of the power model is the component's power consumption. The power model performs the following computation

$$Power = tc(queue\_x_1(0), queue\_x_1(1)) * Coeff_1$$
$$+ \cdots + tc(queue\_x_N(0), queue\_x_N(1)) * Coeff_N$$

where, $tc$ represents the transition count (XOR) function. The inputs to $tc$ come from a set of internal queues that maintain the previous and current values of each component input/output. Since the transition count is a binary value, the multiplications in the power model equation are simply implemented using vector AND gates. The products of the coefficients and respective transition counts are added to obtain the power consumed by the component in the current strobe period.

A circuit description that has been enhanced with power estimation hardware lends itself to simulation using any HDL simulator [2], or emulation using a hardware prototyping platform as suggested in this work.

### 2.2 Design Flow

Figure 2 illustrates the overall flow for power emulation. Given an input RTL design, step **1** maps the RTL components to power models and inserts the corresponding estimation code to create an enhanced RTL description. This description can be fed into any FPGA synthesis, place and route tool flow in step **2**. Finally, the generated netlist can be downloaded to the FPGA emulation platform. As in functional emulation, the testbench can be executed within a simulator, or it can be mapped to the FPGA platform along with the design itself. The outputs of the power aggregator and power models can be observed during emulation to obtain the power consumption in the circuit or any part thereof.

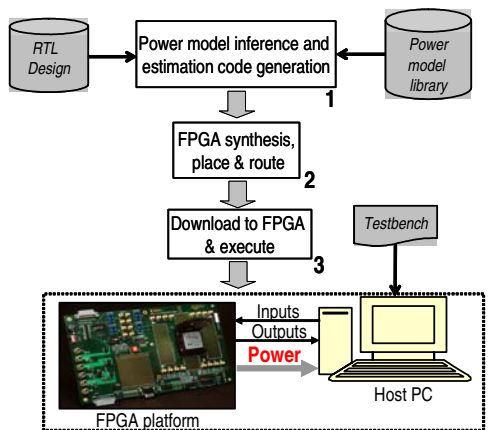

Figure 2: Power emulation flow

## 3 EXPERIMENTAL RESULTS

We present the results of applying the proposed power emulation technique to several industrial designs. In our experimental setup, we begin with a C behavioral description of a design and run a commercial behavioral synthesis tool [10] to synthesize an RTL description. We perform RTL power estimation using (a) PowerTheater [1], and (b) an internal power estimation tool developed at NEC [2]. RTL power estimates and power emulation results are obtained for NEC's `CB130M` [11] $0.13\mu$ standard cell based technology. The RTL circuits were enhanced for power emulation as described in the previous section. Synthesis was performed with Synplify Pro [12], and physical design was performed using tools from Xilinx [13]. For each design, an estimate of power emulation time was computed by measuring the time required to simulate the testbench using the Modelsim HDL simulator and the time to run the design on a PC-based emulation platform using the Xilinx Virtex-II FPGA [13].

The graph of Figure 3 compares the execution times of RTL power estimation (using PowerTheater and NEC's RTL power estimator) with the time required to perform power emulation, for several benchmark designs. The largest design, *MPEG*4, is an MPEG4 decoder used in mobile handset applications. We also report results for the following sub-designs of *MPEG*4: *IDCT* (Inverse Discrete Cosine Transform), *Ispq* (inverse quantization block), and *Vld* (variable length decoder). Other designs include *Bubble_Sort* (sorting circuit), *HVPeakF* (peaking image filter), and *DCT* (Discrete Cosine Transform). The three bars for each benchmark in the graph represent execution times for power estimation using NEC's RTL power estimator (NEC-RTpower), PowerTheater, and power emulation, respectively. The figure also plots the speedup for power emulation compared to the two RTL power estimation tools, with values ranging from $10X$ to over $500X$.

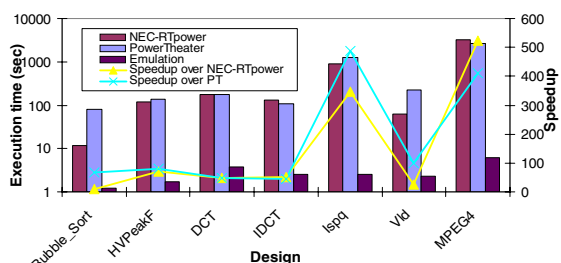

Figure 3: Execution time and speedup for power emulation over RTL power estimation

The above results clearly demonstrate that power emulation has a potential to enable efficient power estimation for large designs. We are presently investigating the challenges that arise in mapping power model enhanced circuits to FPGA prototyping platforms, primarily due to the capacity constraints of FPGA devices. For large designs, we believe that significant work remains to be done in addressing the area occupied by the power estimation hardware before this paradigm can be embodied in a practical methodology for power estimation.